\documentclass[12pt,preprint]{aastex}

\def\Lstar{L_{b_J}^\ast}
\def\Msun{M_\odot}
\def\sigmav{\sigma_v}

\slugcomment{Submitted to {\it The Astrophysical Journal}}

\begin{document}

\title{Concordance between the Dynamics of Satellite Galaxies \\
    in the 2dFGRS and $\Lambda$CDM}

\author{Tereasa G. Brainerd}
\affil{Institute for Astrophysical Research, Boston University,
    725 Commonwealth Ave., Boston, MA 02215}

\begin{abstract}
We compute the velocity dispersion profile, $\sigmav (r_p)$, for the
satellites of host galaxies in the Two Degree Field Galaxy Redshift
Survey (2dFGRS) and in the $\Lambda$CDM GIF simulation.  The host--satellite
selection algorithm yields 1345 host galaxies in the 2dFGRS with luminosities
in the range $0.5~ \Lstar \le L \le 5.5~ \Lstar$, for which a total
of 2475 satellite galaxies is found.  The magnitudes of the galaxies
in the GIF simulation are converted to the $b_J$ band pass, and hosts and 
satellites are selected in the same manner as in the 2dFGRS.  On average,
$\sim 1200$ hosts and $\sim 4100$ satellites are found in the GIF simulation,
where the precise number depends upon the angle from which the simulation is
viewed.  Overall, there is excellent agreement between $\sigmav (r_p)$ for the
satellites in the 2dFGRS and the GIF simulation.  On large
scales, the velocity dispersion
profiles for the complete samples decrease with projected radius,
in good agreement with the
expectations of a CDM universe.  Further, there is a marked dependence of
the velocity dispersion profile on both the host spectral type and the host
luminosity.  In particular, $\sigmav (r_p)$ has a substantially higher
amplitude and steeper slope for satellites of early--type hosts than
it does for satellites of late--type hosts.  In addition, both the
amplitude and slope of $\sigmav (r_p)$ increase with host luminosity.
The velocity dispersion of satellites located within small projected
radii from the host ($r_p \le 120$~kpc) is only marginally consistent with
the local B--band Tully--Fisher relation ($\sigmav \propto L^{0.3}$)
and is fitted
best by a relationship of the form $\sigma_v \propto L^{0.45 \pm 0.10}$.
\end{abstract}


\section{Introduction}

Although it is generally accepted that large, bright galaxies reside
within massive halos of dark matter,  the total mass, the radial
dependence of  the density profile and the physical extent
of  the halos remain poorly--constrained.
A good amount of progress
has been made, however, by the most recent investigations  of weak galaxy--galaxy
lensing  (e.g., Fisher et al.\ 2000; Wilson et al.\ 2001;
McKay et al. 2001; Smith et al.\ 2001;
Guzik \& Seljak 2002;
Hoekstra et al. 2003; Kleinheinrich
et al. 2003, 2004; Hoekstra, Yee \& Gladders 2004) and this method certainly
holds great promise for placing  strong constraints on the nature of dark
matter halos as a function  of cosmic time.  Ideally of course one would hope
to determine the differences between  the physical characteristics of
the halos surrounding galaxies of differing Hubble type
and differing luminosity, and
weak lensing has begun to provide some interesting results regarding
these characteristics.
In particular,
all studies of galaxy--galaxy lensing as a function of the lens  Hubble type
conclude that the velocity dispersion of the halos of
early--type $L^\ast$ galaxies and/or the mass--to--light ratios
of early--type $L^\ast$ galaxies exceed those of late--type
$L^\ast$  galaxies (e.g., Griffiths et al.\ 1996;
McKay et al.\ 2001; Guzik \& Seljak 2002;
Kleinheinrich et al.\ 2003, 2004). 

Although it is abundantly clear that galaxy--galaxy lensing is a very
good tool for probing the halos of galaxies at large physical radii
($ r \gtrsim 100 h^{-1}$~kpc), the  use  of this technique  to constrain
dark matter halos as  a function of galaxy morphology or
galaxy luminosity is complicated by the fact that galaxy--galaxy
lensing is inherently a multiple--deflection problem.  This was first
pointed out by Brainerd, Blandford \& Smail (1996), hereafter BBS,
 in the context of
their analysis of the first statistically--significant ($4\sigma$)
detection of galaxy--galaxy lensing.  In  particular, BBS
found that
more than 50\% of their source galaxies (apparent
magnitudes of $23 < r < 24$) should have been lensed at a
comparable and significant level by {\bf two or  more foreground
galaxies} (apparent magnitudes of $20 < r < 23$).  That is,
for a given source galaxy, it was
clear from the BBS data that the closest lens on the sky to any
given source was
not necessarily the only lens, and neither was it necessarily
the strongest lens.  (See, e.g.,  \S3.6 of BBS.)
Further work by Brainerd (2003, 2004) in an analysis of
multiple weak deflections by the galaxies in the northern Hubble Deep Field 
has  shown that the probability of multiple deflections of
magnitude  $\gamma = 0.005$ (a ``significant'' shear in the context
of galaxy--galaxy lensing) exceeds
50\% for source galaxies  with $z_s \gtrsim 1$ when the  median lens
redshift  is $z_l \sim  0.6$.   
Compared to the case in which only single deflections by the
closest lens galaxy are included in the calculation,
multiple weak deflections
in galaxy--galaxy lensing result in a substantially
higher net tangential shear  about
the lens centers  and
significantly correlated image ellipticities  of the foreground
and background galaxies on angular scales $\theta \lesssim 60''$.
(See also Guzik \& Seljak 2002 for a discussion of the contribution
of group-- and cluster--sized mass distributions to observations
of galaxy--galaxy lensing at lower redshifts.)

As a result of multiple weak deflections, then, computing the
net weak lensing shear due to a particular population of lens galaxies
(i.e., early--type vs.\ late--type; high--luminosity vs.\ 
low--luminosity)
is not simple.  That is, one cannot merely compute the tangential shear
of source galaxies relative to, say, lens galaxies with late--type morphology
and be certain that the shear is caused solely by single deflections
due to lenses with late--type morphology.  This is especially true
in deep data sets (e.g., those for which the median lens redshift
is $\gtrsim 0.5$). Instead, it is
important to analyze and  model the weak  lensing signal in such a  way
that it accounts for
the  fact that multiple deflections are likely to have occurred in the
data.  That is not to say that weak lensing cannot provide
reasonable constraints on the  physical differences between halos
surrounding galaxies of differing morphology  or luminosity; rather,
one  simply has to be careful in  how one arrives  at those constraints.
                                                                                
This in  mind, it is certainly worthwhile to  consider the use of
complementary methods to constrain the nature of dark matter  halos,
and one
such method is, of course, the dynamics of satellite galaxies.  Pioneering
work by Zaritsky \& White (1994) and Zaritsky et al.\ (1997) led to
the  conclusion that  the halos of isolated  spiral galaxies extend
well beyond the optical radii and are extremely massive:
$M(r \lesssim 150h^{-1}~{\rm kpc}) \sim 1 - 2 \times 10^{12} h^{-1}~\Msun$.
Despite these
conscientious analyses, however, a certain skepticism regarding the
usefulness of  this technique remained owing to several important
facts. First, the orbital timescales of the satellites is large (of order
5 to 20 Gyr at sufficiently large radii) and, therefore, virialization
is  by no means guaranteed. Second, there is  no unique host/satellite
selection algorithm which  will
insure the absence of ``interlopers'' from the data (i.e.,
galaxies which are  falsely selected as satellites).  Finally, the sample
sizes in mid--1990's were quite small (of order 70 hosts and 100 satellites),
so the results were based on rather small number statistics.
                                                                                
In the past couple of years, however, the prospects for using 
the dynamics of satellite galaxies to study the dark matter halos of host
galaxies has improved significantly.   
The advent of large redshift surveys such as the Sloan Digital
Sky Survey (SDSS, York et al.\ 2000) and the Two Degree Field 
Galaxy Redshift Survey
(2dFGRS; Colless et al.\ 2001, 2003) has increased the available
samples of potential hosts and satellites by more than an order of
magnitude.  In addition, McKay et al. (2002) and Prada et al. (2003)
have used data from both the SDSS and numerical simulations  to  show  that
the  effects  of interloper galaxies on the inferred velocity dispersion
of  satellite galaxies can be corrected straightforwardly 
by fitting a Gaussian plus a
constant offset to the distribution of velocity differences, $|dv|$,
between the hosts and satellites (see, e.g., Figs.\ 5 and 6 of 
Prada et al.\ 2003).  
Further, Prada et al. (2003)
investigated a number of different algorithms for selecting hosts and satellites
in the SDSS data, including  the algorithm used by McKay et al. (2002),
and found that the results of their dynamical analyses were quite
insensitive to the details of the host--satellite selection algorithm.

Simulations of galaxy redshift surveys by van den Bosch et al.\ (2004) 
seem to
show that the fraction of interloper galaxies is substantially
higher for host--satellite
pairs with small values of $|dv|$ than it is for host--satellite pairs
with large values of $|dv|$, an effect not taken into account by the
recent investigations of satellite dynamics in large surveys.  However,
van den Bosch et al.\ (2004) also note that the inferred value of the
velocity dispersion is very insensitive to the fraction of interloper
galaxies and, hence, recently published values of $\sigmav$ based on
a simple Gaussian--plus--offset fit are unlikely to be substantially
in error. 

In  their analysis of the dynamics of satellite galaxies in
the SDSS, McKay et al. (2002) found that the velocity
dispersion of the satellites was  independent of projected radius
on the sky; i.e., their results were  consistent with isothermal halos.
However, a subsequent analysis of a larger subset of the
SDSS data by Prada et al. (2003)  showed
that when the contamination of the velocity dispersion  due  to
interlopers was expressly calculated as a function of projected
radius (i.e., the fraction of interlopers increases with $r_p$),
the velocity dispersion for the satellites surrounding the SDSS
galaxies decreases with projected radius. 
This is in good agreement
with the expectations for Navarro, Frenk \& White (NFW) halos
(e.g.,  Navarro, Frenk \& White 1997, 1996, 1995), which are thought
to be the most likely halo mass distribution in the context of 
hierarchical structure formation.
                                                                                
Brainerd \&  Specian (2003)  used the 100K data release
of the 2dFGRS to investigate the dynamics of satellite galaxies and,
like Prada  et  al. (2003),
Brainerd  \& Specian (2003) allowed for the fact that the interloper
fraction was  a  strongly increasing  function  of projected  radius.
Unlike Prada  et  al. (2003), however, they found that the velocity dispersion
of the satellites in the 2dFGRS was independent of projected  
radius and, therefore,
that  the halos of the 2dFGRS hosts were consistent with isothermal halos,
not NFW halos.  However,
due to the relatively larger velocity  errors in the 2dFGRS 
($\sigma_{cz} \sim 85$~km~sec$^{-1}$ in the
2dFGRS  vs.\ $\sigma_{cz} \sim 20$~km~sec$^{-1}$
in the SDSS) as well as
the smaller
sample size ($\sim 800$ hosts and 1550 satellites in the 100K  2dFGRS
data vs.\  $\sim 1100$ hosts  and  2700 satellites  in  the  SDSS),
it is entirely possible that the apparent  disagreement over the
radial dependence of the velocity dispersion is due solely
to the larger error bars in the Brainerd \& Specian (2003) analysis.

Even more recently, Conroy et al.\ (2004) investigated the velocity
dispersion profile yielded by 75 satellites surrounding 61 host galaxies
in the DEEP2 redshift survey.  Conroy et al.\ (2004) find that the
velocity dispersion profile of an NFW halo with virial mass
$M_{200} = 5.5\times 10^{12}~h^{-1} \Msun$ is consistent with their
measurements of $\sigmav (r_p)$.  However, a flat velocity dispersion
profile (i.e., a singular isothermal sphere halo) is also formally 
consistent with their data and error bars.
                                                                                
In this paper we continue the efforts of  Brainerd \& Specian
(2003)  and investigate the dynamics of satellites  in the final
data release of  the 2dFGRS.  We compute the radial dependence of the
satellite velocity dispersion as a function of both the host spectral type
and the host luminosity.
In addition, we compare the results from the 2dFGRS galaxies to those obtained
by analyzing the dynamics of satellite galaxies in
the present--epoch  galaxy catalogs of the  flat,
$\Lambda$--dominated GIF  simulation  (Kauffmann et  al.\ 1999).  This is
a publicly--available  simulation which includes  semi--analytic
galaxy formation in a cold dark  matter (CDM)
universe.  The paper  is organized  as follows.  The selection of
hosts and satellites is discussed in \S2. The computation of the
satellite velocity dispersion, $\sigmav (r_p)$, and the correction of
$\sigmav (r_p)$ for velocity errors is discussed in \S3.  Results are shown
in \S4, and a discussion of our results, including a
comparison with previous work, is presented
in \S5.

\section{Selection of Hosts and Satellites}
                                     
\subsection{2dFGRS Galaxies}
                                           
The 2dFGRS is a spectroscopic survey in which the target objects
were selected in the $b_J$ band from the  Automated Plate Measuring
(APM) galaxy survey (Maddox et al.\ 1990a,  1990b) and extensions  to
the original survey.  The  final data release occurred on June 30, 2003
(Colless et al. 2003) and  includes redshifts of 221,414  galaxies
brighter  than $b_J =19.45$  over $\sim 1500$  square  degrees.  All
data, including spectroscopic and photometric catalogs, are
publicly--available from the  2dFGRS website
(\url{http://msowww.anu.edu/au/2dFGRS}), as well as DVDs  that can be  ordered
from the  2dFGRS team.  The  photometric transformation from the
SDSS band passes to $b_J$ is
                                                                                
\begin{equation}
b_J = g' + 0.155 + 0.152(g' -  r')
\label{transform}
\end{equation}
                                                                                
\noindent
(Norberg et al.\ 2002)  and  the absolute magnitude of an
$\Lstar$ galaxy is given by
                                                                                
\begin{mathletters}
\begin{eqnarray}
M_{b_J}^\ast & = & -19.58 \pm 0.05 + 5 \log_{10} h,  \;\;\;\;\;\;\; 
 \eta < -1.4  \\
M_{b_J}^\ast & = & -19.53 \pm 0.03 + 5 \log_{10} h,  \;\;\;\;\;\;\; 
 -1.4 \le \eta < 1.1  \\
M_{b_J}^\ast & = & -19.17 \pm 0.04 + 5 \log_{10} h,  \;\;\;\;\;\;\; 
 1.1 \le \eta < 3.5  \\
M_{b_J}^\ast & = & -19.15 \pm 0.05 + 5 \log_{10} h,  \;\;\;\;\;\;\; 
 \eta \ge 3.5 
\end{eqnarray}
\end{mathletters}
                                                                                
\noindent
where $\eta$ is the spectral  type  of the galaxy (Madgwick et  al.\
2002).  Galaxies with large negative values of $\eta$
have spectra that  are  dominated
by absorption features, and those  with large positive values  of
$\eta$ have spectra  that are dominated by emission lines.  Here,
and throughout this  paper, we  adopt the following values of
the cosmological parameters:  $\Omega_0 = 0.3$,
$\Lambda_0= 0.7$, $H_0 = 70$~km~sec$^{-1}$~Mpc$^{-1}$.
                                                                                
We select a  preliminary set of hosts and satellites from the  2dFGRS
using criteria identical to those of McKay et al.\ (2002), Brainerd
\& Specian (2003), and Sample 3 of Prada et al.\ (2003):
                                                                                
\begin{enumerate}
\item Host galaxies must be ``isolated''.  They must be at least twice
as luminous  as  any other galaxy that  falls within a projected
radius of 2 $h^{-1}$~Mpc and  a  velocity difference of
$|dv| \le 1000$~km~sec$^{-1}$.
\item Potential satellites  must  be at least  4 times fainter than
their host, must fall  within a projected  radius of $500 h^{-1}$~kpc,
and the velocity difference between the host and the satellite
must be $|dv| \le 1000$~km~sec$^{-1}$.
\end{enumerate}

In addition to the above criteria,  we  impose an additional
restriction that the sum total  of
the luminosities  of the satellites  must be less  than the  luminosity
of the host.  This was also done by McKay et al.\ (2002), Prada
et al.\ (2003),  and Brainerd \& Specian (2003) in order to eliminate a
handful of hosts for which the number of satellites is extremely large
and, hence, objects  which are more likely to be in a cluster
environment rather than being  truly isolated.  Further, we eliminate
a small  number of
hosts for which the eyeball morphology provided by the 2dFGRS team falls into the
interaction/merger category, on the grounds that these are dynamically
young systems which are unlikely to be virialized. 
Also, since we will ultimately
be interested in investigating satellite dynamics as a function of
host spectral type, we eliminate a small number of hosts for
which no spectral classification parameter, $\eta$, was  provided
by the 2dFGRS team.   Finally, we restrict our
analysis  to hosts with luminosities in the range
$0.5~ \Lstar \le L \le 5.5~ \Lstar$ since there are relatively few hosts
with $L < 0.5~ \Lstar$ and the distribution of the host--satellite
velocity differences
for hosts with $L >> 5~ \Lstar$ is poorly--fitted by
the technique which we adopt (see below).  This leaves us  with
a final sample that consists of 1345 hosts and 2475 satellites.
The median redshift of the hosts is $z = 0.08$.
                                                                                
The normalized probability  distribution of the 2dFGRS host luminosities
is shown in the left panel of Fig.\ 1, and the normalized probability
distribution of the number  of satellites per host is shown in the
left panel of Fig.\ 2.  The 2dFGRS sample is clearly dominated by
systems containing 1 or 2  satellites per host, with only $\sim 20$\%
of hosts having 3 or more satellites.

\subsection{GIF Galaxies}
 
In order to compare our results  for the 2dFGRS hosts to that expected
for  large, bright galaxies  in a
flat, $\Lambda$-dominated
CDM universe, we have used one of
 the publicly--available GIF simulations to
select samples of theoretical hosts and satellites.  The entire suite
of GIF  simulations
consists of N--body, adaptive P$^3$M simulations
of various CDM universes,  coupled with
a semi--analytic prescription for galaxy formation (see, e.g.,
Kauffmann et al.\ 1999).  Here we use only the GIF simulation
with $\Omega_0 = 0.3$  and $\Lambda_0= 0.7$,  for which the
box length was $141.2 h^{-1}$~Mpc (comoving) and the  particle mass
was $1.4\times 10^{10}  h^{-1}$~Mpc.
                                                                                
Galaxy, halo,
and particle files are all easily downloaded from the GIF project
website, \url{http://www.mpa-garching.mpg.de/GIF}, for a wide range
of redshifts.  Here we make use of only the present--epoch ($z=0$)
data, and the specific galaxy catalog that provides magnitudes  in
the SDSS band passes.  The SDSS magnitudes of the GIF galaxies
were converted to equivalent $b_J$ magnitudes using the transformation
given by equation (\ref{transform}) and, consistent with our adopted 
cosmological parameters,
the absolute $b_J$ magnitudes of the GIF galaxies
were determined using the luminosity function
of Norberg et al.\ (2002):
                                                                                
\begin{equation}
M_{b_J}^\ast - 5 \log_{10} h = -19.66 \pm 0.07 .
\end{equation}
                                                                                
Hosts and satellites in the GIF simulation were selected by
rotating
the simulation randomly and projecting the galaxy distribution along
the line  of sight.  In order to mimic the 2dFGRS data set more
closely, and to test our prescription for accounting for the velocity
errors in the 2dFGRS data, Gaussian--distributed errors with
$\sigma_{cz} = 85$~km~sec$^{-1}$  were added to the line of sight velocities
of the GIF galaxies.  Different velocity errors were assigned
to each galaxy for each rotation of the simulation box.
(The velocity errors
 will, of course,  not only affect the
measured velocity dispersion of the satellites, but they will
also affect the ultimate
selection of hosts and satellites from the galaxy catalog.)
After the addition of the velocity errors,
the host--satellite selection criteria which were
applied to the 2dFGRS data were then applied to the GIF galaxies.
For each rotation of the simulation, the number of hosts and
satellites was similar to that of the 2dFGRS data: $\sim 1200$
hosts and $\sim 4100$ satellites on average.
A total of 100 random rotations of the simulation box were 
performed, and the results shown in all figures correspond to the
mean over these 100 rotations.

The distribution of host luminosities of the GIF galaxies is
shown in the right panel of Fig.\ 1 where, as with the 2dFGRS hosts,
we have restricted the sample to those hosts with $0.5~ \Lstar \le
L \le 5.5~ \Lstar$.  While the  luminosity distribution is fairly
similar for the 2dFGRS and GIF hosts, the GIF hosts are somewhat
more luminous than the 2dFGRS hosts ($L_{\rm med} = 2.3~ \Lstar$ for
the 2dFGRS hosts; $L_{\rm med} = 2.7~ \Lstar$ for the GIF hosts).
The right panel of Fig.\ 2 shows the distribution
of the number of satellites per host in the GIF simulation and, like
the 2dFGRS galaxies, the sample consists primarily of hosts that
have only 1 or 2 satellites, although there are
certainly a 
larger percentage of GIF
hosts with 3 or more satellites.  On the whole, however, the host
and satellite samples in the 2dFGRS and the GIF simulation  are
quite well--matched.
                                                                                
\section{Computation of $\sigmav (r_p)$ and Correction for Velocity Errors}
                                                                                
The velocity dispersion of the satellite galaxies was computed using
the  method championed by McKay et al.\ (2002) and Prada et al.\
(2003).  The distribution of the
observed velocity differences between the hosts
and satellites, $P(|dv|)$, for satellites with projected radii
$r_{1} < r_p \le r_{2}$ is modeled as the sum of a Gaussian
distribution and a constant offset that accounts for the presence
of interlopers in the satellite sample.  For all of the 
host--satellite samples
considered here, this method works well and yields typical values
of $\chi^2$ per degree of freedom in the
range $0.7 \lesssim \chi^2/\nu \lesssim 1.0$.

Similar to the results of van den Bosch et al.\ (2004),
our own study of the GIF simulation suggests that, indeed, 
the fraction of interlopers for host--satellite pairs with small 
values of $|dv|$ exceeds that for host--satellite pairs with large
values of $|dv|$. However, the effect is significantly smaller in the GIF 
simulation than was reported by van den Bosch et al.\ (2004).   In 
addition, like van den Bosch et al.\ (2004)
we find that, for a given distribution of velocity
differences, $P(|dv|)$, the inferred velocity dispersion is not
terribly sensitive to the interloper fraction.  In particular, for
a given $P(|dv|)$, varying the interloper fraction from $f_i \sim 0.1$
to $f_i \sim 0.4$ results in a change in the value
of $\sigmav$ obtained from a Gaussian--plus--offset fit
that is substantially less than the formal error bars
on $\sigmav$.  We therefore conclude that, to within our formal
error bars, the Gaussian--plus--offset fit to $P(|dv|)$ is sufficient
to determine reasonable estimates of $\sigmav$.
                                                                                
Shown in Fig.\ 3 is the velocity dispersion of the satellites
in the GIF simulation, measured as a function of projected radius
from the host.  The solid triangles show the results for a ``raw''
measurement of $\sigmav (r_p)$ in which the velocity errors have been
included in the determination of 
$P(|dv|)$.  The open circles show the results for a measurement of
$\sigmav (r_p)$ in which no velocity errors were assigned to either
the hosts or the satellites.  As expected, the  inclusion of velocity
errors inflates the measured velocity dispersion of the satellites
above what would be obtained in the absence of velocity errors.  Naively,
of course, one would expect the velocity errors to add in quadrature
with the true velocity dispersion
and, hence, in a given radial bin
the velocity dispersion which would be observed in the presence
of the errors  should be:
                                                                                
\begin{equation}
\sigmav^{\rm  obs} = \sqrt{ \left( \sigmav^{\rm true} \right)^2 +
2 \left( \sigma_{cz} \right)^2 },
\label{verr}
\end{equation}
                                                                                
\noindent
where $\sigma_{cz}$ is the typical error in the line of sight velocity
for a single galaxy.
However,
since the hosts and satellites are selected in part on the
basis of velocity differences, the velocity errors will also enter
in to the very definition of  the sample itself and, in  principle,
a naive correction of the velocity dispersion using the above
relation might not be sufficient.
                                                                                
The solid squares in Fig.\ 3, however, show that the naive correction
to the velocity dispersion is sufficient.  That is, the solid squares show
the results of applying equation (\ref{verr})
above to the solid triangles, where we can see a good
agreement between the corrected velocity dispersion profile and the velocity
dispersion profile that was computed in the absence of velocity errors.  We
will, therefore, use equation (\ref{verr}) to correct the measured velocity
dispersions of the satellites in both the 2dFGRS and GIF data, where a
value of $\sigma_{cz} = 85$~km~sec$^{-1}$ is adopted.

\section{Results}
                                                                                
\subsection{Interloper Fraction}

As discussed by Prada et al.\ (2003) and Brainerd \& Specian (2003),
the fraction of interlopers is a strong function of projected radius.
This is due to a geometrical effect; since the volume being searched for
satellites increases with projected radius from the host, the number of
interlopers will necessarily increase with projected radius.
                                                                                
Shown in Fig.\ 4 is the interloper fraction for the 2dFGRS
sample, as well as the interloper fraction obtained for the GIF samples
in  which errors were added to the line of sight velocities.  
Again, the interloper fraction is determined by the constant offset
component that is included when modeling the observed distribution of
velocity differences as a Gaussian plus an offset.  
As with the
luminosity distribution of the hosts (Fig.\ 1) and the distribution of the
number  of satellites (Fig.\ 2), we can see that the 2dFGRS and GIF samples
seem to be well--matched to each other and, hence, a direct
comparison of the results from each would seem to be
justified.

\subsection{Velocity Dispersion Profile for Complete Sample}

Shown in Fig.\ 5 is the radial dependence of the
velocity dispersion of the satellites in the
full 2dFGRS sample, as well as the velocity dispersion of satellites
in the GIF simulation for which velocity errors were added.
In  both cases, the measured velocity dispersions have been corrected
for the velocity errors using equation (\ref{verr}) above with $\sigma_{cz} =
85$~km~sec$^{-1}$ and the interloper fraction was allowed to
vary with projected radius (i.e., Fig.\ 4).
Clearly, there is a
very good agreement between the velocity dispersion of satellite
galaxies in the 2dFGRS and the predictions of a $\Lambda$--dominated
CDM universe.  The velocity dispersion decreases with
radius, similar to the results of Prada et al.\ (2003) for the 
SDSS galaxies,   and shows that, as anticipated, the apparent disagreement
between Brainerd \& Specian (2003) and Prada et al.\ (2003) over the
radial dependence of $\sigmav$ is due to the substantially larger 
error bars in the Brainerd \& Specian (2003) analysis.

A close examination of the velocity dispersion profiles on scales
$\lesssim 300h^{-1}$~kpc in Fig.\ 5 shows that $\sigmav (r_p)$ for the satellites
in the GIF simulation has a somewhat higher amplitude and a somewhat
steeper slope than that for the 2dFGRS satellites.  This is likely due
to the fact that, while the host samples are similar in the
two data sets, they are not identical.
In particular, the median luminosity of GIF hosts is larger than
that of the 2dFGRS hosts by $\sim 0.4~\Lstar$.  Based on known scaling
relations of the internal velocity dispersion (or circular velocity) with
galaxy luminosity, we
certainly anticipate that the velocity dispersion
of satellites of intrinsically bright
galaxies will be larger than that of satellites of intrinsically faint 
galaxies.  We will revisit this in \S4.4 below.

Given the overall good agreement between $\sigmav (r_p)$ for the 2dFGRS and GIF
galaxies, it is not unreasonable to extend our analysis further and
to investigate the dependence of
the satellite velocity dispersion profile
on host spectral type and host luminosity.
The signal--to--noise will necessarily be lower when we subdivide the 
samples but, based on the strength of the signal in Fig.\ 5, it should
be possible to place some modest constraints on the differences between
the dynamics of satellites of hosts with differing spectral types and 
differing luminosities.

\subsection{Dependence of Velocity Dispersion Profile on Host Spectral Type}

The  distribution  of host spectral types in the 2dFGRS host  sample
is shown  in the left panel of
Fig.\ 6, where the parameter $\eta$ is  defined in
Madgwick et al.\ (2002).  Approximately  one third of the hosts have
$\eta \le -2.45$, one third of the hosts have $ -2.45 < \eta < -1.1$,
and one third of the hosts have $\eta  \ge  -1.1$.  From Fig.\ 4 of
Madgwick et al.\ (2002), the  morphologies of these host galaxies
should be approximately E/S0 ($\eta \le -2.45$), Sa ($ -2.45 < \eta
< -1.1$), and Sb/Scd ($\eta \ge -1.1$).  The luminosity distributions
of the hosts within these subsamples is similar, with median
luminosities of $2.64~ \Lstar$ ($\eta \le -2.45$),
$2.25~ \Lstar$ ($-2.45 < \eta < -1.1$), and $2.11~ \Lstar$
($\eta \ge -1.1$).  We also note that, although the 2dFGRS team has
provided eyeball morphologies for some of the galaxies, these are 
restricted to only the brightest hosts in our sample ($b_J \lesssim 18$)
and choosing to subdivide the sample based on $\eta$ rather than 
eyeball morphology will allow for the largest possible subsamples
of hosts.  In addition, a cursory examination of a subset of the 
hosts that have 2dFGRS eyeball classifications of ``spiral''  
shows that some of these objects have spectra that are inconsistent
with spiral morphology (i.e., their spectra are strongly dominated by absorption
lines) and, so, the reliability of the eyeball classification 
seems to be somewhat questionable.
            
As above, we  compute the velocity dispersions of the satellites
about the hosts, allowing
for the  fact that the interloper fraction  will  increase with
projected radius.  The results are shown in Fig.\ 7, where it
is clear that the velocity dispersion profiles of the satellites
of early--type
hosts are significantly different from those of late--type
hosts.  That is, while in all three cases 
the velocity dispersion profiles 
are decreasing,  $\sigmav (r_p)$ has a much higher
amplitude and steeper decline for the satellites of early--type
hosts than it does for the satellites of 
late--type hosts.  Although there is some
difference in the median luminosities of hosts with different
values of $\eta$, we will show below that they are not sufficiently
different for the trends in  $\sigmav (r_p)$  in Fig.\ 7 to be caused 
primarily
by the differences in host luminosity.  In other words, the differences
in the three panels of Fig.\ 7 are most strongly correlated 
the spectral type of the host, not its luminosity. 
                                                                                
Differences in the velocity dispersion profiles for the halos of
elliptical galaxies and spiral galaxies are, of course, expected
at some level  due to the fact that the ellipticals are very likely
to be merger products.  While we cannot compare the predictions
of the GIF simulation directly to the results of the 2dFGRS on the basis of the
spectral parameter, we can at least compare the results for
GIF hosts of differing color.  Shown in the right
panel of Fig.\ 6, then, is the
distribution of $(g'-r')$ colors for the hosts in the GIF simulation.
The distribution is clearly bi--modal and, so, we investigate the velocity
dispersion of the satellites of ``blue'' GIF hosts, $(g'-r') < 0.2$, and
``red'' GIF hosts, $(g'-r') > 0.2$.  We also note that, although 
a bi--modality in the $(g'-r')$ colors of SDSS galaxies is well--established
(e.g., Baldry et al.\ 2004;  Hogg et al.\ 2004; Blanton et al.\ 2003;
Kauffmann et al.\ 2003), the bi--modality seen in the right panel
of Fig.\ 6 is much sharper than that shown by the SDSS galaxies, and
the median value of $(g'-r')$ for the ``blue'' GIF hosts is 
much bluer than that of
the SDSS galaxies.  This is, of course, simply a reflection of the fact
that while the GIF simulation yields some remarkable agreements with
the known local galaxy populations, it is not perfect in its representation.

The results for the velocity dispersion profiles of
the satellites of blue and red GIF hosts  are shown in Fig.\ 8
and, at least qualitatively, they are in general
agreement with the results for
the 2dFGRS hosts.  That is, $\sigmav (r_p)$ for  the satellites
of the red GIF hosts has a much higher amplitude than that
for the satellites of the blue GIF hosts.  In addition, $\sigma (r_p)$
for the satellites of the red GIF hosts decreases more rapidly with
projected radius than does $\sigmav (r_p)$ for the blue GIF hosts.
However, $\sigmav (r_p)$ for the satellites of the red GIF hosts has
a lower amplitude and a shallower slope than does $\sigmav (r_p)$ for the
satellites of the early--type 2dFGRS hosts (i.e., left panel of Fig.\ 7).
This disagreement persists even if we restrict our analysis to
the very reddest GIF hosts; i.e., we obtain the same velocity
dispersion profile for GIF hosts with $(g' - r') > 0.45$ as
we do for GIF hosts with $(g' - r') > 0.2$.  Similar to
the satellites of the red GIF hosts, $\sigmav (r_p)$ for the satellites
of the blue GIF hosts has a lower amplitude than $\sigmav (r_p)$ for
the late--type 2dFGRS hosts (i.e., middle and right panels of Fig.\ 7).
In addition,
the slope of $\sigmav (r_p)$ for the satellites of the blue GIF
hosts is consistent with zero, while $\sigmav (r_p)$ decreases 
for the satellites of the late--type 2dFGRS hosts.

\subsection{Dependence of Velocity Dispersion Profile on Host Luminosity}

On the basis of the Tully--Fisher and Faber--Jackson
relations, we expect the velocity dispersions of the satellite galaxies to
be strongly correlated with the luminosities of the host galaxies.
Shown in Fig.\ 9, then, are the velocity dispersion profiles for the satellites
of four independent subsamples of the hosts, where the median host luminosity
in the subsamples is $\Lstar$ (top left panel), $2~ \Lstar$ (top right
panel), $3~ \Lstar$ (bottom right panel), and $4 ~\Lstar$ (bottom
right panel).  Solid squares show the results for satellites of the
GIF hosts, and open circles with error bars show the results for
satellites of the 2dFRGS hosts.  Although the signal--to--noise is
somewhat low for $\sigmav (r_p)$ for the satellites of the 2dFGRS galaxies,
overall there is excellent agreement between the results for the
2dFGRS galaxies and the GIF galaxies.  In addition, close examination
of Fig.\ 9 shows that as the median luminosity of the host increases,
both the amplitude and the slope of $\sigmav (r_p)$ increase.  Given the 
good agreement between the 2dFGRS and GIF galaxies in each of the
individual panels of Fig.\ 9, then, the small differences in
$\sigmav (r_p)$ that are seen 
on scales $\lesssim 300h^{-1}$~kpc in Fig.\ 5 are most likely due to
the fact that the distribution of host luminosities is somewhat
different in the GIF simulation than it is in the 2dFGRS.

Finally, in Fig.\ 10 we show the dependence of the velocity dispersion
of the satellites as a function of host luminosity for small 
projected radii ($r_p \le 84 h^{-1}$~kpc; i.e., $r_p \le 120$~kpc for our
adopted value of $H_0$).  The choice of this particular
physical scale is motivated
by making a direct comparison to the results of Prada et al.\ (2003),
who computed $\sigmav$  for the satellites
of SDSS galaxies as a function of host absolute magnitude for
projected radii $r_p \le 120$~kpc.  As with Fig.\ 9, we again see a
very good agreement between the results for the satellites of the
2dFGRS galaxies and the GIF galaxies.  The relationship between
$\sigmav$ and host luminosity is well--fitted by a power law of
index 0.45, where the error on the power law index is 0.10 for
the 2dFGRS satellites.  This is somewhat steeper than one
would expect on the basis of the local B-band Tully--Fisher
relation, $\sigmav \propto L^{0.3}$ (Verheijen 2001), but is only 
discrepant at less than the $2\sigma$ level.

\section{Discussion}

Here we have investigated the radial dependence of the velocity
dispersion of satellite galaxies about host galaxies in an observational
sample (the 2dFGRS) and a theoretical sample (the $\Lambda$CDM GIF 
simulation).  Hosts and satellites were selected from the observational
and theoretical samples using identical criteria, and velocity errors
which are comparable to the velocity errors in the observational sample
were added to the line of sight velocities in the theoretical sample.  Overall,
there is a remarkable similarity between the observational and theoretical
samples of hosts and satellites (e.g., similar number of hosts and
satellites, similar host luminosity distribution, similar number of
satellites per host, and a similar fraction of interlopers).

We have shown that a simple correction for the errors
that were added to the line of sight velocities of the
GIF galaxies successfully reproduces the velocity dispersion profile
that is measured in the absence of these errors.  Further,
when we correct the measured velocity dispersion profile of the 
full sample of 2dFGRS satellites using the same simple correction, 
we find a very good agreement of  $\sigmav (r_p)$ for the 2dFGRS 
satellites and $\sigmav (r_p)$ for the GIF satellites.  
Further, $\sigmav (r_p)$ decreases with $r_p$ and on
the large physical scales investigated here, this decrease is
consistent with the expectations of NFW halos.

We have investigated the dependence of $\sigmav (r_p)$ on the
host spectral type, $\eta$, in the 2dFGRS and find a clear difference
between the velocity dispersion profiles of the satellites of early--type
hosts and late--type hosts.  Although all of the 
velocity dispersion profiles decrease with $r_p$, both the
amplitude and the slope of $\sigmav (r_p)$ for the satellites of early--type
hosts are substantially larger than those for the satellites of
late--type hosts.  This general trend is also shown by the satellites
of the GIF hosts, where both the slope and the amplitude of $\sigmav (r_p)$ for
the satellites of the red GIF hosts are larger than those for the
satellites of the blue GIF hosts.  The differences in $\sigmav (r_p)$
for hosts of differing color are, however, much less marked in
the simulation than they are in the 2dFGRS.

We have also investigated the dependence of $\sigmav (r_p)$ on the
host luminosity, and find good agreement between the satellites of
the 2dFGRS galaxies and the GIF galaxies.  In all cases, $\sigmav (r_p)$
decreases with $r_p$ and both the slope and the amplitude
of $\sigmav (r_p)$ increase with host luminosity.  In addition, the
small--scale ($r_p \le 84h^{-1}$~kpc) velocity dispersion of the satellites
scales identically with host luminosity in the 2dFGRS and GIF samples:
$\sigmav \propto L^{0.45}$.

A velocity dispersion profile that decreases at large
projected radii is, of course, expected in CDM universes.
In addition, based on the
known
scalings of internal velocity dispersion or circular velocity with
the luminosity of the galaxy, we expect the amplitude of $\sigmav (r_p)$
to increase with increasing host luminosity.  Our results for
$\sigmav (r_p)$ as a function of host luminosity (i.e., Fig.\ 9) show
precisely this trend, and are in good agreement with the results
of Prada et al.\ (2003) for the velocity dispersion profiles of the
satellites of SDSS galaxies.  The most direct comparison between
our results and those of Prada et al.\ (2003) is the upper right
panel of Fig.\ 9 in this paper and Fig.\ 7 of Prada et al.\ (2003).
In this case our hosts have $L_{\rm med} = 2 \Lstar$ and the hosts
in Prada et al.\ (2003) have comparable luminosities ($-20.5 < M_B 
< -21.5$).  Comparing these two figures, we find an excellent agreement
of both the slope and the amplitude of $\sigmav (r_p)$ in these independent
analyses.  Fig.\ 8 of Prada et al.\ (2003) shows $\sigmav (r_p)$ for
fainter SDSS hosts ($-19.5 < M_B < -20.5$), and shows a clear decrease
in the amplitude compared to their results for brighter hosts.  Unfortunately,
the faintest luminosity subsample in our work (top left panel of
Fig.\ 9 in this paper) is somewhat too bright
to be compared directly to the faintest subsample in Prada et al.\ (2003).
As result, although we do see a clear decrease in the amplitude of
$\sigmav (r_p)$ in going from the top right panel to the top left panel of
our Fig.\ 9, the decrease is not as substantial as seen when comparing
Figs.\ 7 and 8 of Prada et al.\ (2003).  
Finally, the slope of $\sigmav (r_p)$ in the Prada et al.\ (2003)
analysis does not appear to increase
as a function of host luminosity; however, the increase in the
slope of $\sigmav (r)$ with host luminosity in our Fig.\ 9
is sufficiently gradual that it likely would not have been obvious
for the range of host luminosities shown in Figs.\ 7 and 8 of 
Prada et al.\ (2003).

The dependence of the small--scale ($r_p \le 84h^{-1}$~kpc) satellite
velocity dispersion with host luminosity is, however, only marginally
consistent between this work and that of Prada et al.\ (2003), who found 
$\sigmav \propto L^{0.3}$ for the same physical scale.   The explanation
for this is unclear, but given the wide range of published constraints
on halo velocity dispersions from both dynamical and weak lensing
studies, this apparent discrepancy is perhaps not unexpected.  
Based on satellite dynamics, the
constraints on the value of the index of the Tully--Fisher relation range from
approximately zero (i.e., Zaritsky et al.\ 1997) to 1 (Brainerd \& Specian
2003), with rather large error bars.  Complicating matters further, these
dynamical
constraints come from different physical scales and the only results that
can truly be compared directly are those of this work and Prada et al.\
(2003).  Similarly, weak lensing constraints on the value of the index
of the Tully--Fisher index range from 0.3 to 1.7 (e.g., Hudson et al.\
1998;  McKay et al.\ 2001; Smith et al.\ 2001; Kleinheinrich et al.\ 2003, 2004).
The weak lensing measurement that claims the most accurate value,
$0.30^{+0.16}_{-0.12}$ (Kleinheinrich et al.\ 2004), is in
modest agreement with our value but, again, the weak lensing measurement
is made over a somewhat larger physical scale ($\lesssim 150h^{-1}$~kpc)
and, moreover, was determined for lens galaxies with much higher
redshifts than our host galaxies.

Finally, this work and Brainerd \& Specian (2003) are, so far, the only 
investigations of satellite dynamics to be performed
as a function of the host spectral
type or morphology.  Brainerd \& Specian (2003) used the eyeball 
morphologies provided by the 2dFGRS team for the 100K data release to
divide their sample into 159 hosts with elliptical or S0 morphologies
and 243 hosts with spiral morphologies.  Unlike this work,
Brainerd \& Specian (2003) found
that $\sigmav (r_p)$ was independent of $r_p$ for the satellites
of both the early--
and late--type hosts; however, the error bars in
Brainerd \& Specian (2003) were larger due to the relatively small
sample sizes and, so, a decrease in $\sigmav (r_p)$ at the level
that we see here would not have been
detected in their data.  
In addition, our recent examination of the spectra of 2dFGRS
hosts with eyeball classifications of ``spiral'' suggest that at least
some have been misclassified (c.f.\ \S4.3 above) and, so, a detailed 
comparison of this investigation and that of Brainerd \& Specian (2003)
is somewhat questionable.  

However, our observed general trend for the velocity dispersion of the 
satellites of early--type hosts to be greater than that of the
satellites of late--type hosts  is in agreement with recent
weak lensing constraints on $\sigmav$ for the halos of galaxies with
differing morphology.  In particular, Kleinheinrich et al.\ (2003)
find that, averaged over projected radii of
$20 h^{-1}~{\rm kpc} \le r_p \le 150h^{-1}~{\rm kpc}$, the
velocity dispersion of $L^\ast$ early--type lenses is
$198^{+32}_{-42}$~km sec$^{-1}$, while for $L^\ast$ late--type lenses
it is $146^{+32}_{-38}$~km sec$^{-1}$.  From Fig.\ 7, the velocity 
dispersion of the satellites in the 2dFGRS averaged over a similar
scale is $\sim 290$~km sec$^{-1}$ for hosts with 
$\eta \le -2.45$, $\sim 230$~km sec$^{-1}$ for hosts with 
$-2.45 < \eta < -1.1$, and $\sim 160$~km sec$^{-1}$ for hosts with
$\eta \ge -1.1$.  The median luminosities of the 2dFGRS hosts are, of
course, brighter than $\Lstar$ so we would expect the mean
velocity dispersions of the satellites to exceed the values found
by Kleinheinrich et al.\ (2003).  Based on the scaling of $\sigmav$
with host luminosity that we found in Fig.\ 10, we would expect
the velocity dispersion of early--type lenses with $L = 2L^\ast$ in
the Kleinheinrich et al.\ (2003) sample to
be of order 300~km sec$^{-1}$ and the velocity dispersion of
late--type lenses with $L = 2L^\ast$ to be of order 200~km sec$^{-1}$.
Both of these are in quite reasonable agreement with our results for
the 2dFGRS satellites as a function of host spectral type, where
the late--type hosts are expected to be those with $\eta > -2.45$.

We conclude that the dynamics of the satellite galaxies in the
2dFGRS are in good agreement with the expectations for satellite
galaxies in a $\Lambda$CDM universe, and that the available samples
of hosts and satellites in current redshift surveys are now becoming
sufficiently large that they can be used to study 
the dark matter halos of the hosts.  Given the potential complications
in the interpretation of galaxy--galaxy lensing data
(e.g., multiple weak deflections caused by foreground galaxies, as
well as galaxy lenses being embedded within groups and clusters),
the use of the dynamics of 
satellites to study the dark matter halos of large
galaxies is likely to emerge as a useful
technique that is entirely complementary to galaxy--galaxy lensing.

\acknowledgments

It is a pleasure to thank the members of the 2dFRGS team and the 
GIF project for not only making their data publicly--available, but
also for making their data products superbly easy to obtain and use.
Support under NSF contract AST--00984572 is also gratefully acknowledged.


\clearpage
\begin{figure}
\plotone{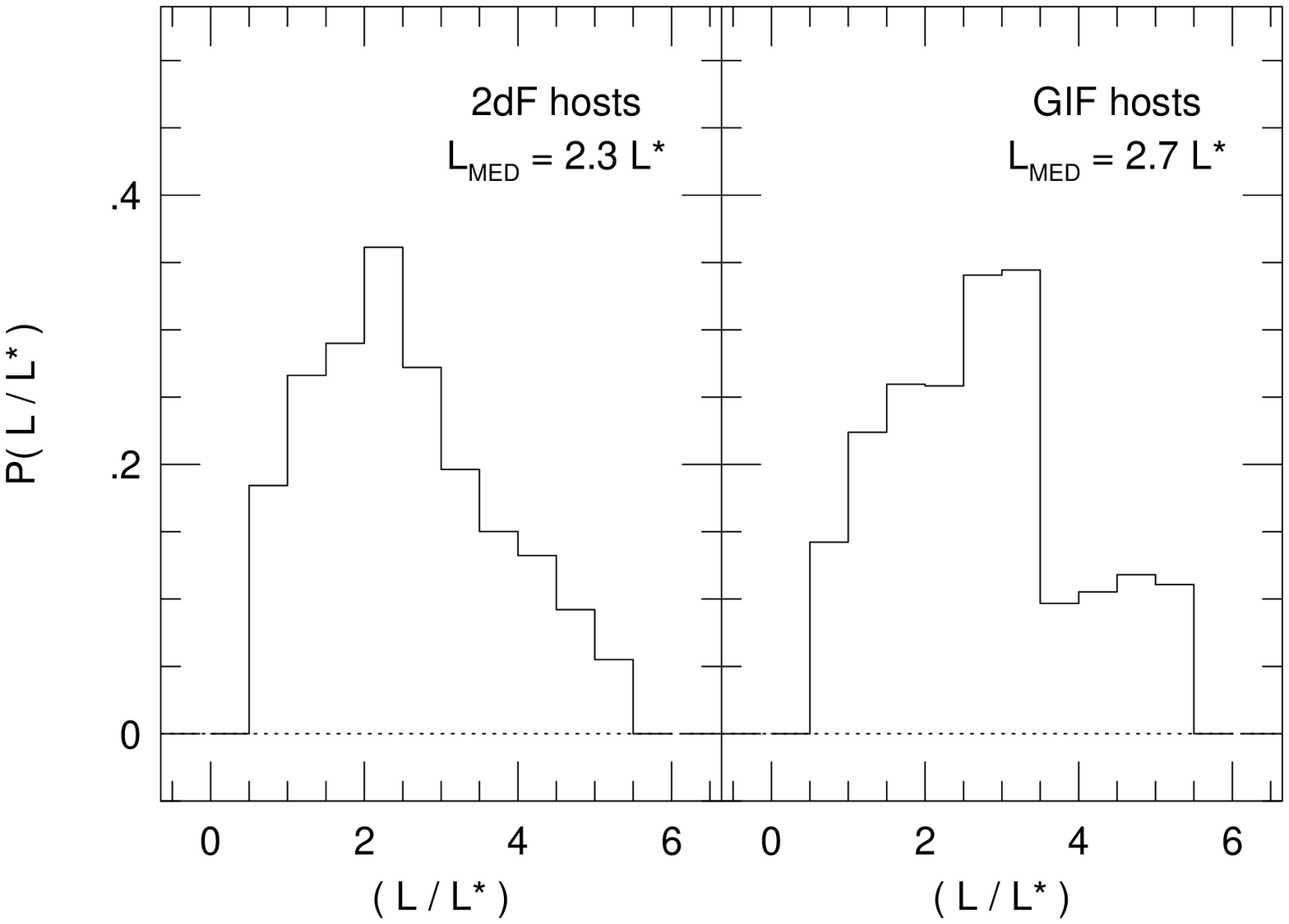}
\vskip -9.0cm
\caption{
Left panel: Probability distribution of host 
galaxy luminosities
in the 2dFGRS.  The sample has been restricted to 
hosts with luminosities in the range $0.5~ \Lstar \le L_{\rm host} \le 
5.5~ \Lstar$.
Right panel: Same as left panel, but for host galaxies in the
GIF simulation.}
\end{figure}

\clearpage
\begin{figure}
\plotone{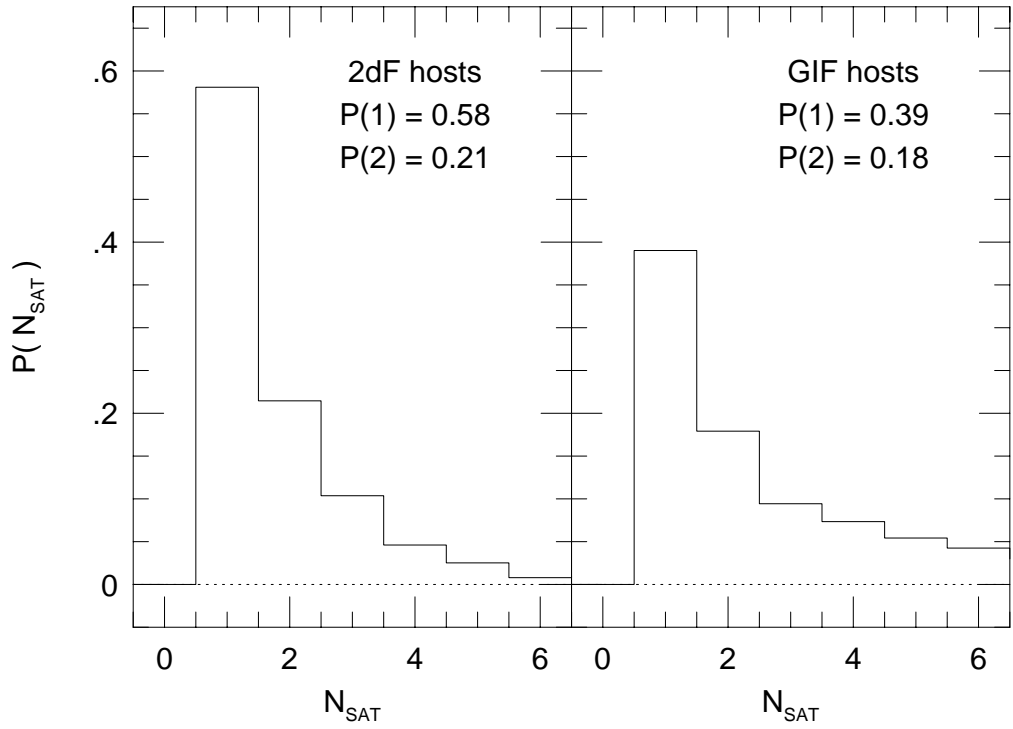}
\vskip -9.0cm
\caption{
Left panel: Probability distribution of the
number of satellite galaxies associated
with host galaxies in the 2dFGRS.  Here, by definition, each host
galaxy has at least one satellite. 
Right panel: Same as left panel, but for host galaxies in the
GIF simulation.}
\end{figure}

\clearpage
\begin{figure}
\plotone{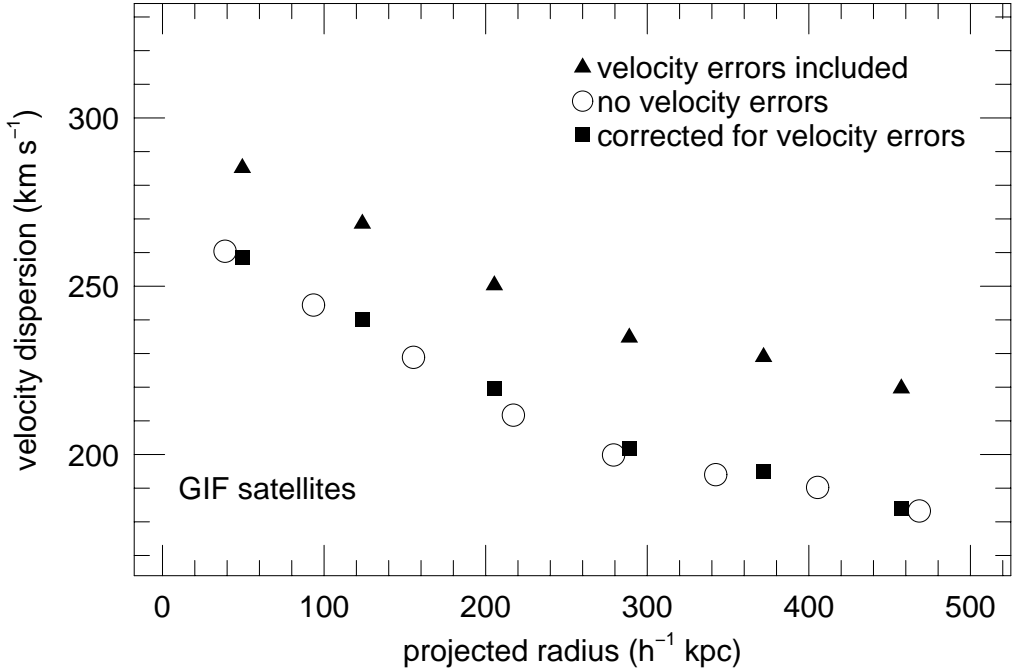}
\vskip -9.0cm
\caption{
Velocity dispersion profile,
$\sigmav (r_p)$, for  satellite galaxies in the GIF simulation.
Here the interloper
fraction has been allowed to vary with projected radius.  Solid triangles
show the raw values $\sigmav (r_p)$ that are
obtained when errors that are comparable
to the errors in the line of sight velocities of the 2dFGRS galaxies are added
to the line of sight velocities of the GIF galaxies.  Open circles show 
$\sigmav (r_p)$ that is obtained when no errors are added to the
velocities of the 
GIF galaxies.  Solid squares show $\sigmav (r_p)$ that is obtained after
correcting the solid triangles for the velocity errors using equation 
(\ref{verr}).
}
\end{figure}

\clearpage
\begin{figure}
\plotone{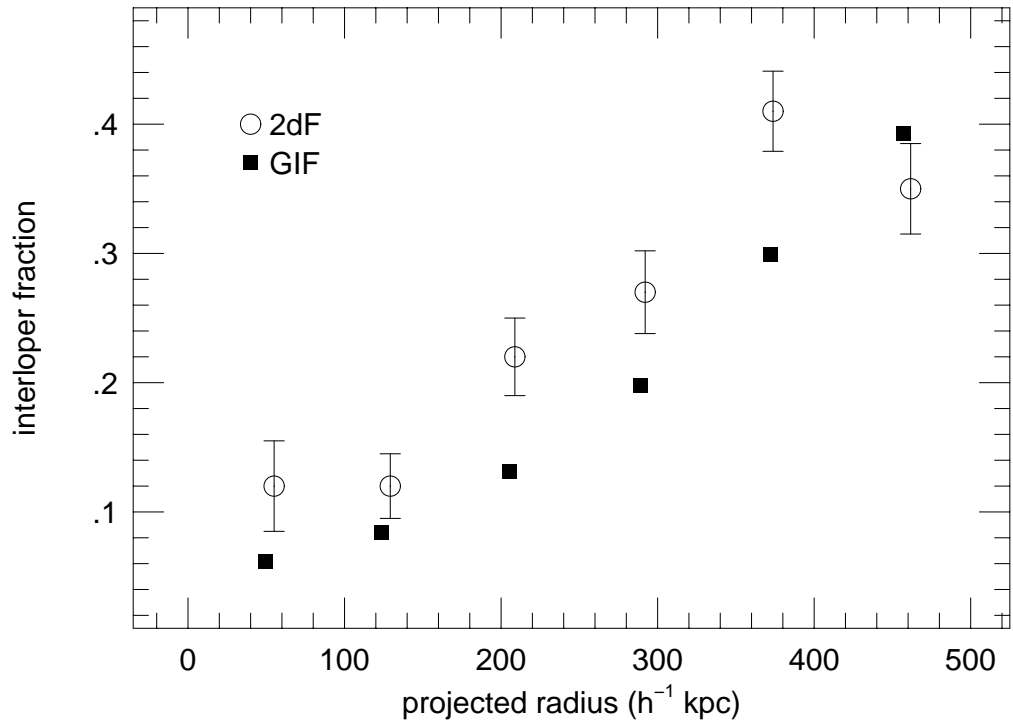}
\vskip -9.0cm
\caption{
Interloper fraction as a function of radius in the 2dFGRS
(open circles) and the GIF simulation (solid squares).
}
\end{figure}

\clearpage
\begin{figure}
\plotone{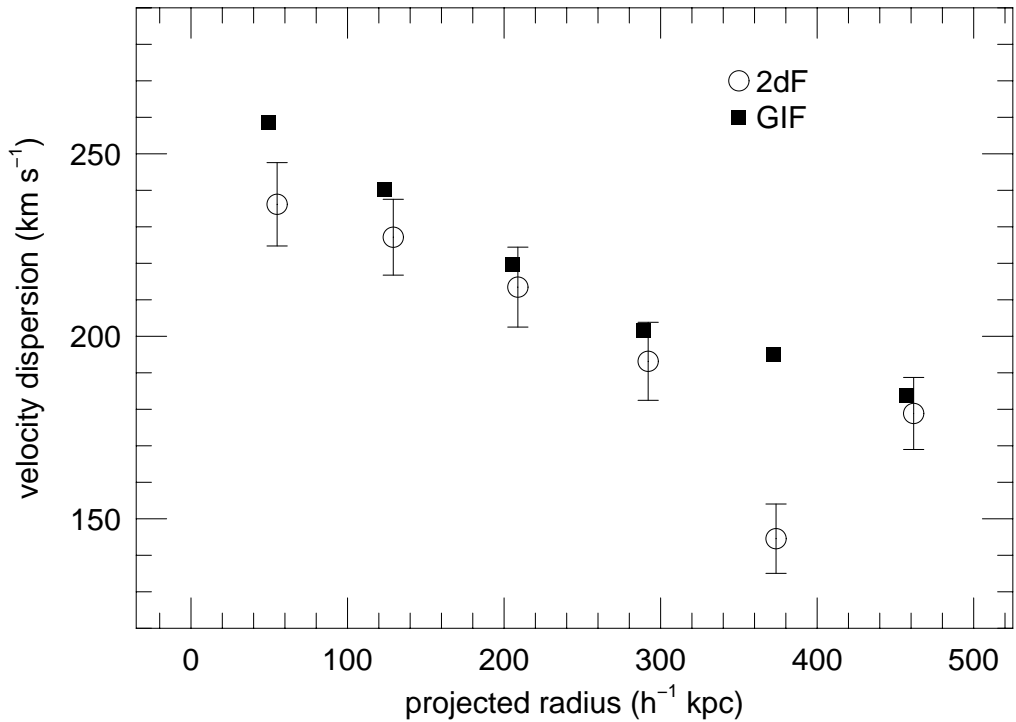}
\vskip -9.0cm
\caption{
Velocity dispersion profiles for satellite galaxies
in the 2dFGRS
(open circles) and the GIF simulation (solid squares). 
Values of $\sigmav (r_p)$ have been corrected for errors in the line of sight
velocity using equation (\ref{verr}).  On scales $\lesssim 300 h^{-1}$~kpc,
the slight difference between  $\sigmav (r_p)$ for 
the 2dFGRS satellites and $\sigmav (r_p)$ for the GIF satellites is likely due
to the difference in median luminosities of the 2dFGRS and GIF
hosts (see \S4.4 and Fig.\ 9).
}
\end{figure}

\clearpage
\begin{figure}
\plotone{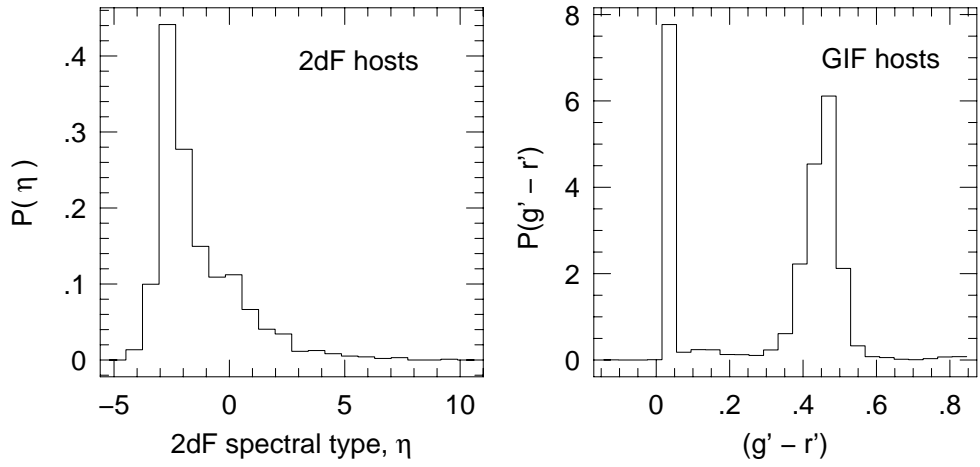}
\vskip -12.0cm
\caption{
Left Panel: Probability distribution of host spectral type in
the 2dFGRS.  Right panel: Probability distribution of host
$(g'-r')$ color in the GIF simulation.
}
\end{figure}

\clearpage
\begin{figure}
\plotone{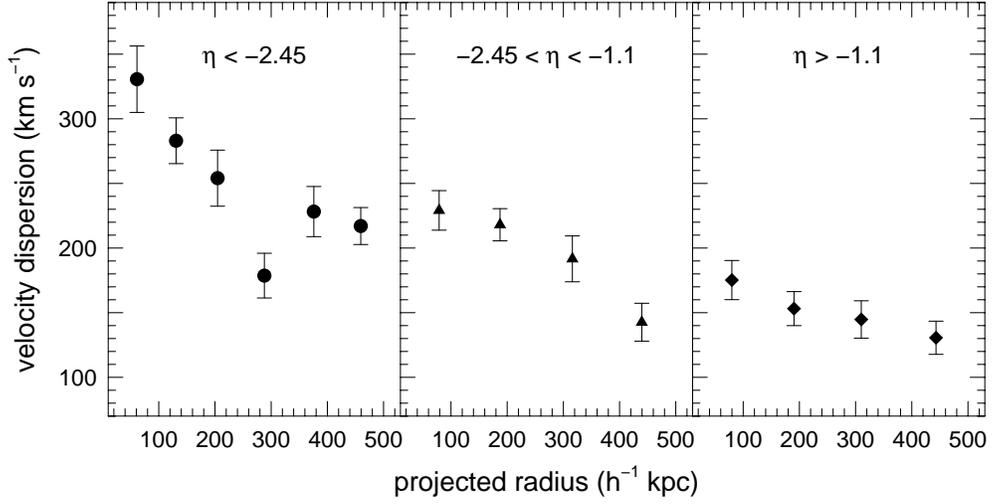}
\vskip -12.0cm
\caption{
Dependence of 2dFGRS satellite velocity dispersion profiles on the
spectral type of the host, $\eta$.
Based on Fig.\ 4 of Madgwick et al.\
(2002), the morphology of the hosts is expected to be roughly
E/S0 in the left panel, Sa in the middle panel, and Sb/Scd in the
right panel.
Values of $\sigmav (r_p)$ have been corrected for errors in the
line of sight velocity using equation (\ref{verr}).
}
\end{figure}

\clearpage
\begin{figure}
\plotone{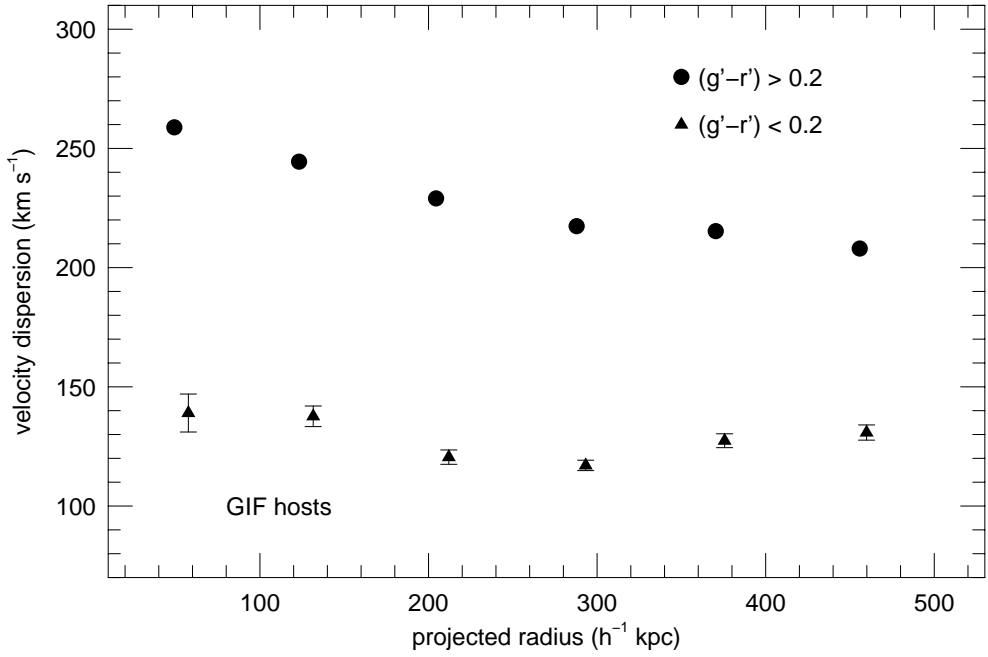}
\vskip -10.0cm
\caption{
Dependence of GIF satellite velocity dispersion profiles on
the $(g'-r')$ color of the host.
Values of $\sigmav (r_p)$ have been corrected for errors in the
line of sight velocity using equation (\ref{verr}).
}
\end{figure}

\clearpage
\begin{figure}
\plotone{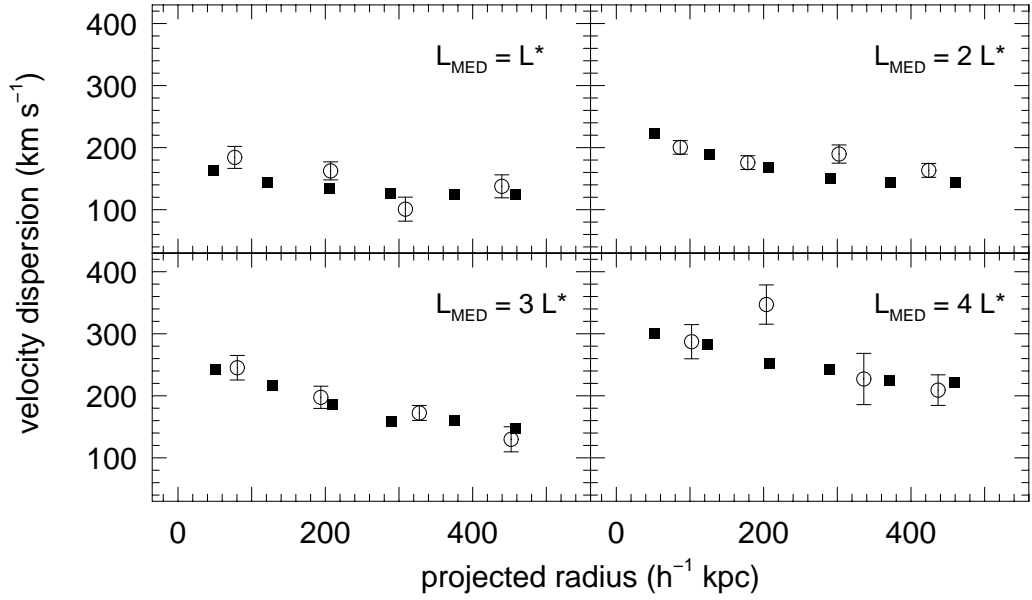}
\vskip -11.0cm
\caption{
Dependence of satellite velocity dispersion profiles on the
luminosity of the host (open circles: 2dFGRS; solid
squares: GIF simulation).
Values of $\sigmav (r_p)$ have been corrected for errors in the
line of sight velocity using equation (\ref{verr}).
}
\end{figure}

\clearpage
\begin{figure}
\plotone{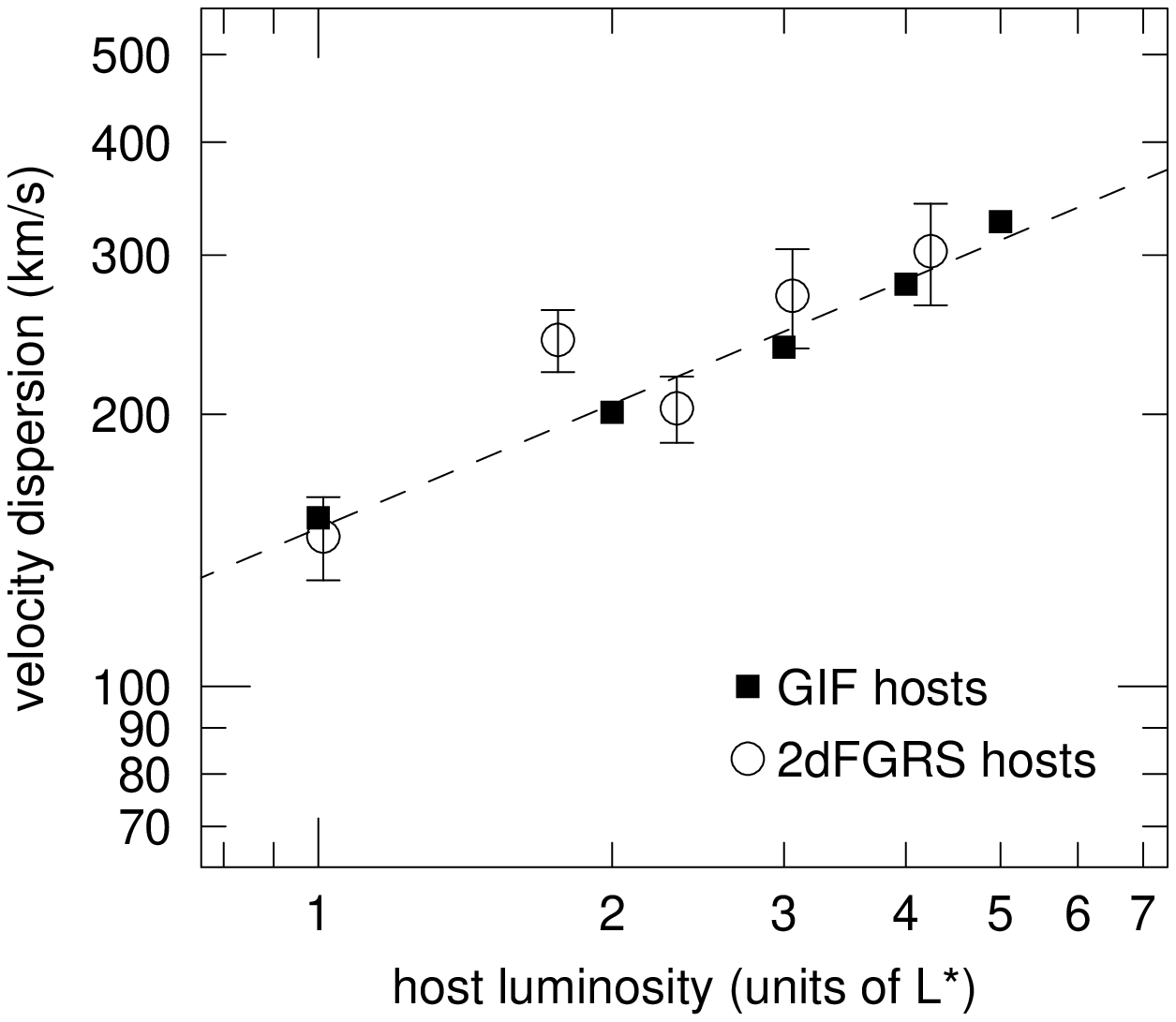}
\vskip -9.0cm
\caption{
Dependence of satellite velocity dispersion on host luminosity
for satellites with projected radii 
$r_p \le 120$~kpc (i.e., $r_p \le 84 h^{-1}$~kpc).  Open
circles show results for the 2dFGRS galaxies; solid squares
show results for the GIF galaxies.  Dashed line shows
$\sigmav \propto L^{0.45}$, which is the best--fit to both the
2dFGRS galaxies and the GIF galaxies.
Values of $\sigmav$ have been corrected for errors in the
line of sight velocity using equation (\ref{verr}).
}
\end{figure}


\begin{thebibliography}{}

\bibitem{} Baldry, I. K., Glazebrook, K., Brinkmann, J., Ivezic, Z.,
Lupton, R. H., Nichol, R. C. \& Szalay, A. S. 2004, ApJ, 600, 681

\bibitem{} Blanton, M. R., Hogg, D. W., Bahcall, N. A., Baldry, I. K.,
Brinkmann, J., Csabai, I, Eisenstein, D., Fukugita, M., Gunn, J. E., 
et al.\ 2003, ApJ, 594, 186

\bibitem{} Brainerd, T. G. 2003, in proceedings of ``Hubble's Science
Legacy: Future Optical--UV Astronomy from Space'', ASP Conf. Series
vol.\ 291, eds.\ Sembach, K. R., Blades, J. C., Illingworth, G. D.
\& Kennicutt, R. C., 347

\bibitem{} Brainerd, T. G. 2004,
to be published in
the proceedings of IAU Symposium 225, ``The Impact of Gravitational
Lensing on Cosmology'', eds.\ Y. Mellier \& G. Meylan, astro--ph/0409374

\bibitem{} Brainerd, T. G., Blandford, R. D. \& Smail, I. 1996,
ApJ, 466, 623 (BBS)

\bibitem{} Brainerd, T. G. \& Specian, M. A. 2003, ApJ, 593, L7

\bibitem{} Colless, M., Dalton, G., Maddox, S., Sutherland, W., 
Norberg, P., Cole, S., Bland--Hawthorn, J., et al.\ 2001, MNRAS,
328, 1039

\bibitem{} Colless, M. Peterson, B. A., Jackson, C., Peacock, J. A.,
Cole, S., Norberg, P., Baldry, I. K. et al.\ 2003 (astro--ph/0306581)

\bibitem{} Conroy, C., Newman, J. A., Davis, M., Coil, A. L., Yan, R.,
Cooper, M. C., Gerke, B. F., Faber, S. M. \& Koo, D. 2004
(astro--ph/0409305)

\bibitem{} Fischer, P., McKay, T. A., Sheldon, E., Connolly, A.,
Stebbins, A., Frieman, J. A., Jain, B., et al.\ 2001, AJ, 120, 1198

\bibitem{} Griffiths, R. E., Casertano, S., Im, M. \& Ratnatunga, K.
1996, MNRAS, 282, 1159

\bibitem{} Guzik, J. \& Seljak, U. 2002, MNRAS, 335, 311

\bibitem{} Hoekstra, H., Franx, M., Kuijken, K., Carlberg, R. G. \&
Yee, H. K. C. 2003, MNRAS, 340, 609

\bibitem{} Hoekstra, H., Yee, H. K. C. \& Gladders, M. D. 2004,
ApJ, 606, 67

\bibitem{} Hogg, D. W., Blanton, M. R., Brinchmann, J., Eisenstein, D. J.,
Schlegel, D. J., Gunn, J. E., McKay, T. A., Bahcall, N. A., Brinkmann, J. \&
Meiksin, A. 2004, ApJ, 601, L29

\bibitem{} Maddox, S. J., Efstathiou, G., Sutherland, W. J. \& Loveday, J. 1990a,
MNRAS, 243, 692
                                                                                
\bibitem{} Maddox, S. J., Efstathiou, G., Sutherland, W. J. \& Loveday, J. 1990b,
MNRAS, 246, 433

\bibitem{} Madgwick, D. S., Lahav, O., Baldry, I. K., Baugh, C. M., 
Bland--Hawthorn, J., Bridges, T., Cannon, R., Cole, S., Colless, M., 
et al.\ 2002, MNRAS, 333, 133

\bibitem{} Norberg, P., Cole, S., Baugh, C. M., Frenk, C. S., Baldry, I.,
Bland--Hawthorn, J., Bridges, T., Cannon, R., Colless, M., et al.\ 2002,
MNRAS, 336, 907

\bibitem{} Navarro, J.F., Frenk, C.S. \& White, S.D.M. 1995,
MNRAS, 275, 720

\bibitem{} Navarro, J.F., Frenk, C.S. \& White, S.D.M. 1996,
ApJ, 462, 563

\bibitem{} Navarro, J.F., Frenk, C.S. \& White, S.D.M. 1997,
ApJ, 490, 493

\bibitem{} Kauffmann, G., Heckman, T. M., White, S. D. M., Charlot, S.,
Tremonti, C., Brinchmann, J., Bruzual, G., Peng, E. W., Seibert, M., et al.\
MNRAS, 341, 33

\bibitem{} Kauffmann, G., Colberg, J. M., Diaferio, A. \& White,
S. D. M. 1999, MNRAS, 303, 188

\bibitem{} Kleinheinrich, M., Schneider, P., Erben, T., Schirmer, M.,
Rix, H.--W. \& Meisenheimer, K. 
2003, to be published in the proceedings of ``Gravitational Lensing:
A Unique Tool for Cosmology''
(astro--ph/0304208)

\bibitem{} Kleinheinrich, M., Rix, H.--W., Schneider, P., Erben, T.,
Meisenheimer, K., Wolf, C. \& Schirmer, M. 2004, to be published in
the proceedings of IAU Symposium 225, ``The Impact of Gravitational
Lensing on Cosmology'', eds.\ Y. Mellier \& G. Meylan (astro--ph/0409320)

\bibitem{} McKay, T. A., Sheldon, E. S., Racusin, J., Fischer, P., Seljak,
U., Stebbins, A., Johnston, D., et al.\ 2001 (astro--ph/0108013)

\bibitem{} McKay, T. A., Sheldon, E. S., Johnston, D., Grebel, E. K.,
Prada, F., Rix, H.--W., Bahcall, N. A., Brinkmann, J., et al.\ 2002,
ApJ, 571, 85

\bibitem{} Prada, F., Vitvitska, M., Klypin, A., Holtzman, J. A.,
Schelgel, D. J., Grebel, E. K., Rix, H.--W., Brinkmann, J.,
McKay, T. A. \& Csabai, I. 2003, ApJ, 598, 260

\bibitem{} Smith, D. R., Bernstein, G. M., Fischer, P. \& Jarvis, M.
2001, ApJ, 551, 643

\bibitem{} Wilson, G., Kaiser, N., Luppino, G. A. \& Cowie, L. L.
2001, ApJ, 555, 572

\bibitem{} van den Bosch, F. C., Norberg P., Mo, H. J. \& Yang, X.
2004 (astro--ph/0404033)

\bibitem{} Verheijen, M. A. W. 2001, ApJ, 563, 694

\bibitem{} York, D. G., Adelman, J., Anderson, J. E., Anderson, S. F.,
Annis, J., Bahcall, N. A., Bakken, J. A., Barkhouser, R., et al.\
2000, AJ, 120, 1579
                                                                                
\bibitem{} Zaritsky, D. \& White, S. D. M. 1994, ApJ, 435, 599
                                                                                
\bibitem{} Zaritsky, D., Smith, R., Frenk, C. \& White, S. D. M.,
1997, ApJ, 478, 39
\end{thebibliography}
\end{document}